\documentstyle{elsart}
\input{epsf}
\newcommand{\SS}{\renewcommand{\arraystretch}{1.2}}
\begin{document}
\begin{frontmatter}

\title{\bf Multidimensional Resonance Analysis of 
$\Lambda_c^+ \rightarrow pK^-\pi^+$}

\author[9]{E.~M.~Aitala,}
       \author[1]{S.~Amato,}
    \author[1]{J.~C.~Anjos,}
    \author[5]{J.~A.~Appel,}
       \author[14]{D.~Ashery,}
       \author[5]{S.~Banerjee,}
       \author[1]{I.~Bediaga,}
       \author[8]{G.~Blaylock,}
    \author[15]{S.~B.~Bracker,}
    \author[13]{P.~R.~Burchat,}
    \author[6]{R.~A.~Burnstein,}
       \author[5]{T.~Carter,}
 \author[1]{H.~S.~Carvalho,}
    \author[12]{N.~K.~Copty,}
    \author[9]{L.~M.~Cremaldi,}
 \author[18]{C.~Darling,}
       \author[5]{K.~Denisenko,}
       \author[3]{S.~Devmal,}
       \author[11]{A.~Fernandez,}
       \author[12]{G. F. Fox,}
       \author[2]{P.~Gagnon,}
       \author[1]{C.~Gobel,}
       \author[9]{K.~Gounder,}
     \author[5]{A.~M.~Halling,}
       \author[4]{G.~Herrera,}
 \author[14]{G.~Hurvits,}
       \author[5]{C.~James,}
    \author[6]{P.~A.~Kasper,}
       \author[5]{S.~Kwan,}
    \author[12]{D.~C.~Langs,}
       \author[2]{J.~Leslie,}
       \author[5]{B.~Lundberg,}
       \author[1]{J. Magnin,}
       \author[14]{S.~MayTal-Beck,}
       \author[3]{B.~Meadows,}
 \author[1]{J.~R.~T.~de~Mello~Neto,}
    \author[7]{D.~Mihalcea,}
    \author[16]{R.~H.~Milburn,}
 \author[1]{J.~M.~de~Miranda,}
       \author[16]{A.~Napier,}
       \author[7]{A.~Nguyen,}
  \author[3,11]{A.~B.~d'Oliveira,}
       \author[2]{K.~O'Shaughnessy,}
    \author[6]{K.~C.~Peng,}
    \author[3]{L.~P.~Perera,}
    \author[12]{M.~V.~Purohit,}
       \author[9]{B.~Quinn,}
       \author[17]{S.~Radeztsky,}
       \author[9]{A.~Rafatian,}
    \author[7]{N.~W.~Reay,}
    \author[9]{J.~J.~Reidy,}
    \author[1]{A.~C.~dos Reis,}
    \author[6]{H.~A.~Rubin,}
    \author[9]{D.~A.~Sanders,}
 \author[3]{A.~K.~S.~Santha,}
 \author[1]{A.~F.~S.~Santoro,}
       \author[3]{A.~J.~Schwartz,}
       \author[4,17]{M.~Sheaff,}
    \author[7]{R.~A.~Sidwell,}
    \author[18]{A.~J.~Slaughter,}
    \author[3]{M.~D.~Sokoloff,}
    \author[1]{J.~Solano,}
       \author[7]{N.~R.~Stanton,}
     \author[5]{R.~J.~Stefanski,}
       \author[17]{K.~Stenson,}                                        
    \author[9]{D.~J.~Summers,}
 \author[18]{S.~Takach,}
       \author[5]{K.~Thorne,}
    \author[7]{A.~K.~Tripathi,}
       \author[17]{S.~Watanabe,}
 \author[14]{R.~Weiss-Babai,}
       \author[10]{J.~Wiener,}
       \author[7]{N.~Witchey,}
       \author[18]{E.~Wolin,}
       \author[7]{S.~M.~Yang,}
       \author[9]{D.~Yi,}
       \author[7]{S.~Yoshida,}
       \author[13]{R.~Zaliznyak,}
       \author[7]{C.~Zhang,}              
                             
\collab{Fermilab E791 Collaboration}            

\address[1]{Centro Brasileiro de Pesquisas F\'\i sicas, Rio de Janeiro RJ, Brazil}

\address[2]{University of California, Santa Cruz, California 95064}              

\address[3]{University of Cincinnati, Cincinnati, Ohio 45221}                    

\address[4]{CINVESTAV, 07000 Mexico City, DF Mexico}    

\address[5]{Fermilab, Batavia, Illinois 60510}                                   

\address[6]{Illinois Institute of Technology, Chicago, Illinois 60616}           

\address[7]{Kansas State University, Manhattan, Kansas 66506}                    

\address[8]{University of Massachusetts, Amherst, Massachusetts 01003}           

\address[9]{University of Mississippi-Oxford, University, Mississippi 38677}            


\address[10]{Princeton University, Princeton, New Jersey 08544}                

\address[11]{Universidad Autonoma de Puebla, Puebla, Mexico}                

\address[12]{University of South Carolina, Columbia, South Carolina 29208}     

\address[13]{Stanford University, Stanford, California 94305}                  

\address[14]{Tel Aviv University, Tel Aviv, 69978 Israel}                   

\address[15]{Box 1290, Enderby, British Columbia, V0E 1V0, Canada}          

\address[16]{Tufts University, Medford, Massachusetts 02155}                   

\address[17]{University of Wisconsin, Madison, Wisconsin 53706}                

\address[18]{Yale University, New Haven, Connecticut 06511}

\date{\today}

\begin{abstract}
We present the results of a five-dimensional resonant amplitude 
analysis of the $\Lambda_c^+{\rightarrow}pK^{-}\pi^{+}$ system based 
on $946\pm 38$ reconstructed decays. These data were  produced in 500 
GeV/$c$ $\pi^-$-N interactions by Fermilab experiment E791. 
We report measurements of the amplitudes for $\Lambda_c^+$ decay into
nonresonant $pK^-\pi^+$ and to $p\overline{K}^{*0}(890)$,
$\Delta^{++}(1232)K^-$, and $\Lambda(1520)\pi^+$ and we comment on other
possible 
resonant enhancements. This is the first complete amplitude analysis of
the $\Lambda_c^{+}{\rightarrow}pK^{-}\pi^{+}$ system. We find that
$(54.8\pm5.5\pm3.5)$\% of the decays are nonresonant,
$(19.5\pm2.6\pm1.8)$\% of the decays are via the $\overline{K}^{*0}$
resonance, $(18.0\pm2.9\pm2.9)$\% of the decays are via the
$\Delta^{++}$ resonance, and $(7.7\pm1.8\pm1.1)$\% of the decays are via
the $\Lambda(1520)$ resonance. We find evidence for an increasingly
negative polarization of the $\Lambda_c^+$ baryons as a function of
$p_T^2$, in agreement with a recent model\cite{dha} and with a related
measurement\cite{jez}. 

\end{abstract}


\end{frontmatter}

\def\Kbar{$\overline{K}$}
\def\ckv{\v Cerenkov\ }
\def\lbrk{\hfil\break}

\section{Introduction}

Since the discovery of charm in 1974, great progress has been made in
the understanding of charm meson decays. Charm meson lifetimes,
branching fractions, and resonant decays have been studied
extensively\cite{pdg}. While charm baryon lifetimes and branching
fractions have also been measured, these measurements are not as
complete, and no information is available on the amplitude analysis of
any charm baryon decay system. Measurements of charm baryon decays yield
information regarding the relative importance of spectator and exchange
amplitudes. The latter are thought to play a leading role in enhancing
charm baryon decay rates.  The most obvious components of the
$\Lambda_c^+\to pK^-\pi^+$ decays (and charge conjugate decays, which
are implied throughout this paper) include the nonresonant $pK^-\pi^+$
decay as well as the $p\overline{K}^{*0}(890)$ and $\Lambda(1520)\pi^+$
two-body decays. All three of these decays can be described by spectator
and W-exchange amplitudes. In lowest order, the $\Delta^{++}(1232)K^-$
decay can occur only via the exchange amplitude. Exchange amplitudes are
suppressed in charm meson decays, at least at the quark level, because
of helicity and form-factor effects.  These effects are not expected to
inhibit exchange amplitudes for charm baryons due to the three-body
nature of the interaction. To understand fully this system of decays, as
well as other charmed baryon decays, a complete resonant amplitude
analysis is needed.

The charm baryon and its decay products carry spin and 
the charm baryon may be polarized upon production. Previous 
charm pseudoscalar meson decay analyses have studied structure in 
the two-dimensional space of the decay
product effective masses (Dalitz plot distributions),
but the spin effects just described require five
kinematic variables for a complete description. While this complicates
the analysis, it affords greater sensitivity to the parameters of
interest. As a by-product of the analysis, the production polarization 
of the $\Lambda_c^+$, ${\bf P}_{\Lambda_c}$, is also measured.
This analysis is the first five-dimensional amplitude analysis
and, as such, is unique. 

\section{Formalism}

We parameterize the observed decay rate as a function of the 
$\Lambda_c^+$ polarization, ${\bf P}_{\Lambda_c}$, 
and of the magnitudes and relative phases of each 
intermediate two-body resonance decay amplitude. We assume that the nonresonant 
decay is described by an amplitude that is constant across phase
space. The differential decay rate $\d\Gamma$ (or signal density $S$) may
be expressed as 

\begin{eqnarray}
\d\Gamma \sim S(\vec x) & = &
\frac{(1 + {\bf P}_{\Lambda_c})}{2}
( | \sum_{\mathrm{r}} B_{\mathrm{r}}(m_{\mathrm{r}})\alpha_{\mathrm{r,\frac12,\frac12}}|^2 
+ |\sum_{\rm{r}} B_{\rm{r}}(m_{\rm{r}})\alpha_{\rm{r,\frac12,-\frac12}}  |^2 )\nonumber\\
 & + & 
\frac{(1 - {\bf P}_{\Lambda_c})}{2} 
( |\sum_{\rm{r}} B_{\rm{r}}(m_{\rm{r}})\alpha_{\rm{r,-\frac12,\frac12}}  |^2
+ |\sum_{\rm{r}} B_{\rm{r}}(m_{\rm{r}})\alpha_{\rm{r,-\frac12,-\frac12}}  |^2 )
\label{eqsigdens}  
\end{eqnarray}
where $\alpha_{\rm{r,m,\lambda_{\rm{p}}}}$ is the complex decay amplitude for 
resonance $\rm{r}$ given $\rm{m}$, the spin projection of the $\Lambda_c$ on 
the z-axis, and $\lambda_{\rm{p}}$, the proton helicity in
the $\Lambda_c$ rest frame. 

$B_{\rm{r}}(m_{\rm{r}})$ in Equation \ref{eqsigdens} is the 
normalized relativistic Breit-Wigner amplitude
corrected for the centrifugal barrier\cite{pd}.
Given the decay mode $\Lambda_c \to r(\to a b) c$, 
\begin{equation}
B_{\rm{r}}(m_{\rm{r}}) = (-2|p_c||p_a|)^L\frac{F_{\Lambda_c}F_{\rm{r}}}{m_0^2-m_{\rm{r}}^2-im_0\Gamma_{\rm{r}}}
\end{equation}
where
\begin{equation}
\Gamma_{\rm{r}} = \Gamma_0(\frac{q}{q_0})^{2L+1}\frac{m_0}{m_{\rm{r}}}\frac{F_{\rm{r}}^2(q)}{F_{\rm{r}}^2(q_0)}
\end{equation}
for resonance $\rm{r}$ of angular momentum $L$ at the reconstructed two body mass $m_{\rm{r}}$
with the momentum $q$ (and $q_0$ when $m_{\rm{r}}$ = $m_0$) 
of a daughter particle in the resonance's rest frame, and 
with resonance mass and width $m_0$ and $\Gamma_0$ as found in Ref. \cite{pdg}.
Using this convention, we set $B_{\rm{r}}(m_{\rm{r}})$ for the 
nonresonant decay to be 1.  $F_L$ is the strong coupling factor at 
the appropriate decay vertex, and takes the Blatt-Weisskopf form as
described in Table \ref{Ftable}. Table \ref{Rtable} lists the
range of the strong interaction, $R_X$.

\vskip0.25in

\begin{table}[htbp]
\caption{The expressions for $F$ used in
         the Breit-Wigner amplitude.\lbrk}
\label{Ftable}
\begin{tabular}{cc}\hline
  $L$ &                           $F_L$ \\ \hline
    0 &                               1 \\ 
    1 &           $(1+R_X^2q^2)^{-1/2}$ \\ 
    2 & $(9+3R_X^2q^2+R_X^4q^4)^{-1/2}$ \\ \hline
\end{tabular}
\end{table}

\vskip0.25in

\begin{table}[htbp]
\renewcommand{\arraystretch}{1.55}
\caption{The values of $R$ used in
         the Breit-Wigner amplitude.\lbrk}
\lbrk
\label{Rtable}
\begin{tabular}{cc}\hline
         $X$             & $R_X$              \\ 
\omit                    & (GeV/$c^2$)$^{-1}$ \\ \hline
$\overline{K}^{*0}(890)$ & 3.4  \cite{aston}  \\ 
$\Delta^{++}(1232)$      & 5.22 \cite{koch}   \\ 
$\Lambda(1520)$          & 6.29 \cite{watson} \\ 
$\Lambda_c^+$            & 5.07 \cite{pilk}   \\ \hline
\end{tabular}
\end{table}

\vskip0.25in

In Tables \ref{amp1}--\ref{amp4}, where the amplitudes  
$\alpha_{\rm{r,m,\lambda_p}}$ (derived using the helicity formalism 
described in Ref. \cite{jawi}) can be seen more explicitly, the direction 
($\phi_{\rm{r}}$,$\theta_{\rm{r}}$) is the direction of the 
resonance, r, in the $\Lambda_c$ rest frame, using the 
convention of Ref. \cite{kor}.  The primed angles
refer to the direction of one of the resonance's daughters in the
resonance's rest frame. Note that the decay amplitudes for each
resonance may have contributions to each of the four terms in Equation 1.

\begin{figure}[htbp]
        \centering \leavevmode
        \epsfxsize=5.4in
        \epsfbox[89 450 550 633]{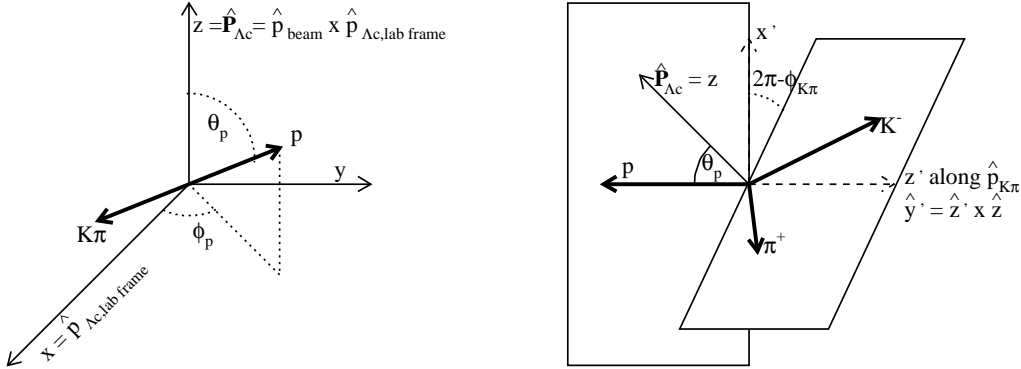}
        \caption{Definition of angles using
$\Lambda_c^+ \rightarrow p\overline{K}^{*0} \rightarrow pK^-\pi^+$
as an example. In both figures the $\Lambda_c^+$ is at rest. In the
first figure, which defines ($\theta_{\rm{p}},\phi_{\rm{p}}$), the $x$-axis
is along the direction of motion of the $\Lambda_c^+$ in the lab frame
and the $z$-axis is the polarization axis, normal to the plane of
production. In the second figure we define $\phi_{K\pi}$ as the
angle between the plane containing the $\overline{K}^{*0}$ decay
products and the plane containing the proton and the $x$-axis.
}\label{angles1222}
\end{figure}

Each event in the final data sample is described
by five kinematic
variables of interest (two two-body masses and the decay angles
$\theta_{\rm{p}}$, $\phi_{\rm{p}}$, and $\phi_{\rm{K\pi}}$ as defined in Figure
\ref{angles1222}) which are determined after 
the pK$\rm{\pi}$ reconstructed mass is 
constrained to the $\Lambda_c$ mass.  We chose the 
quantization axis (the $z$-axis in the $\Lambda_c$ rest frame) to
be normal to the $\Lambda_c$ production plane (as defined by ${\hat
p}_{\rm{beam}} \times {\hat p}_{\Lambda_c}$, where ${\hat
p}_{\rm{beam}}$ is the beam direction and ${\hat p}_{\Lambda_c}$ is
the $\Lambda_c$ production direction in the lab frame). The $x$-axis in
the $\Lambda_c$ rest frame is chosen to be the direction of the
$\Lambda_c$ in the lab frame.

\begin{table}[htbp]
\renewcommand{\arraystretch}{2.95}
\caption{Amplitudes for the $\Lambda_c^+ (\frac12^+) \to  
(\overline{K}^{*0}(890) (1^-) \to K^-\pi^+)  p (\frac12^+)$ decay mode.}
\vskip 5pt
\label{amp1}
\begin{tabular}{ccc}\hline
$\rm{m}$ & $\lambda_{\rm{p}}$ & Amplitude \\ \hline
$\frac12$ &  $\frac12$ & 
	$E_1{\rm e}^{i\phi_{E_1}} 
	\d^{\frac12 }_{\frac12 \frac12}(\theta_{\rm{\overline{K}^{*0}}})
        \d^{1}_{1 0}(\theta'_{\rm{K}}) {\rm e}^{i\phi'_{\rm{K}}} + 
        E_2{\rm e}^{i\phi_{E_2}} 
	\d^{\frac12 }_{\frac12 -\frac12}(\theta_{\rm{\overline{K}^{*0}}})
        \d^{1}_{0 0}(\theta'_{\rm{K}}) 
	{\rm e}^{i\phi_{\rm{\overline{K}^{*0}}}}$ \\ 
$\frac12$ & -$\frac12$ & $E_3{\rm e}^{i\phi_{E_3}} \d^{\frac12 }_{ \frac12  \frac12}(\theta_{\rm{\overline{K}^{*0}}})
               \d^{1    }_{        0       0 }(\theta'_{\rm{K}}) +
                       E_4{\rm e}^{i\phi_{E_4}} \d^{\frac12 }_{ \frac12 -\frac12}(\theta_{\rm{\overline{K}^{*0}}})
               \d^{1    }_{       -1       0 }(\theta'_{\rm{K}})
                                             {\rm e}^{i(\phi_{\rm{\overline{K}^{*0}}} - \phi'_{\rm{K}})}$ \\ 
-$\frac12$ &  $\frac12$ & $E_1{\rm e}^{i\phi_{E_1}} \d^{\frac12 }_{-\frac12  \frac12}(\theta_{\rm{\overline{K}^{*0}}})
               \d^{1    }_{        1       0 }(\theta'_{\rm{K}}) 
                       {\rm e}^{-i(\phi_{\rm{\overline{K}^{*0}}} - \phi'_{\rm{K}})}+
                       E_2{\rm e}^{i\phi_{E_2}} \d^{\frac12 }_{-\frac12 -\frac12}(\theta_{\rm{\overline{K}^{*0}}})
               \d^{1    }_{        0       0 }(\theta'_{\rm{K}})$ \\ 
-$\frac12$ & -$\frac12$ & $E_3{\rm e}^{i\phi_{E_3}} \d^{\frac12 }_{-\frac12  \frac12}(\theta_{\rm{\overline{K}^{*0}}})
               \d^{1    }_{        0       0 }(\theta'_{\rm{K}}) {\rm e}^{-i\phi_{\rm{\overline{K}^{*0}}}} +
                       E_4{\rm e}^{i\phi_{E_4}} \d^{\frac12 }_{-\frac12 -\frac12}(\theta_{\rm{\overline{K}^{*0}}})
               \d^{1    }_{       -1      0 }(\theta'_{\rm{K}}) {\rm e}^{-i\phi'_{\rm{K}}}$ \\ 
\hline
\end{tabular}
\end{table}

\begin{table}[htbp]
\renewcommand{\arraystretch}{2.95}
\vskip 15pt
\caption{Amplitudes for the $\Lambda_c^+ (\frac12^+) \to 
(\Delta^{++}(1232) (\frac32^+) \to p\pi^+) K^-$ decay mode.}
\vskip 5pt
\label{amp2}
\begin{tabular}{ccc}\hline
$\rm{m}$ & $\lambda_{\rm{p}}$ & Amplitude \\ \hline
$\frac12$ &  $\frac12$ & $F_1{\rm e}^{i\phi_{F_1}} \d^{\frac12 }_{ \frac12     \frac12 }(\theta_{\rm{\Delta^{++}}})
               \d^{\frac32 }_{    \frac12   \frac12 }(\theta'_{\rm{p}}) +
                       F_2{\rm e}^{i\phi_{F_2}} \d^{\frac12 }_{ \frac12    -\frac12 }(\theta_{\rm{\Delta^{++}}})
               \d^{\frac32 }_{   -\frac12   \frac12 }(\theta'_{\rm{p}})
                       {\rm e}^{i(\phi_{\rm{\Delta^{++}}} - \phi'_{\rm{p}})}$ \\ 
$\frac12$ & -$\frac12$ & $F_1{\rm e}^{i\phi_{F_1}} \d^{\frac12 }_{ \frac12     \frac12 }(\theta_{\rm{\Delta^{++}}})
               \d^{\frac32 }_{    \frac12  -\frac12 }(\theta'_{\rm{p}}) {\rm e}^{i\phi'_{\rm{p}}} +
                       F_2{\rm e}^{i\phi_{F_2}} \d^{\frac12 }_{ \frac12    -\frac12 }(\theta_{\rm{\Delta^{++}}})
               \d^{\frac32 }_{   -\frac12  -\frac12 }(\theta'_{\rm{p}}) {\rm e}^{i\phi_{\rm{\Delta^{++}}}} $ \\ 
-$\frac12$ &  $\frac12$ & $F_1{\rm e}^{i\phi_{F_1}} \d^{\frac12 }_{-\frac12     \frac12 }(\theta_{\rm{\Delta^{++}}})
               \d^{\frac32 }_{    \frac12   \frac12 }(\theta'_{\rm{p}}) {\rm e}^{-i\phi_{\rm{\Delta^{++}}}} +
                       F_2{\rm e}^{i\phi_{F_2}} \d^{\frac12 }_{-\frac12    -\frac12 }(\theta_{\rm{\Delta^{++}}})
               \d^{\frac32 }_{   -\frac12   \frac12 }(\theta'_{\rm{p}}) {\rm e}^{-i\phi'_{\rm{p}}}$ \\ 
-$\frac12$ & -$\frac12$ & $F_1{\rm e}^{i\phi_{F_1}} \d^{\frac12 }_{-\frac12     \frac12 }(\theta_{\rm{\Delta^{++}}})
               \d^{\frac32 }_{    \frac12  -\frac12 }(\theta'_{\rm{p}}) 
                       {\rm e}^{-i(\phi_{\rm{\Delta^{++}}} - \phi'_{\rm{p}})} +
                       F_2{\rm e}^{i\phi_{F_2}} \d^{\frac12 }_{-\frac12    -\frac12 }(\theta_{\rm{\Delta^{++}}})
               \d^{\frac32 }_{   -\frac12  -\frac12 }(\theta'_{\rm{p}})$ \\ 
\hline
\end{tabular}
\end{table}

\begin{table}[htbp]
\renewcommand{\arraystretch}{2.95}
\vskip 15pt
\caption{Amplitudes for the $\Lambda_c^+ (\frac12^+) \to (\Lambda(1520) 
(\frac32^-) \to pK^-)\pi^+$ decay mode.}
\vskip 5pt
\label{amp3}
\begin{tabular}{ccc}\hline
$\rm{m}$ & $\lambda_{\rm{p}}$ & Amplitude \\ \hline
$\frac12$ &  $\frac12$ & $ H_1{\rm e}^{i\phi_{H_1}} \d^{\frac12 }_{ \frac12 \frac12 }(\theta_{\rm{\Lambda(1520)}})
                \d^{\frac32 }_{    \frac12   \frac12 }(\theta'_{\rm{p}}) +
                       H_2{\rm e}^{i\phi_{H_2}} \d^{\frac12 }_{ \frac12 -\frac12 }(\theta_{\rm{\Lambda(1520)}})
                \d^{\frac32 }_{   -\frac12   \frac12 }(\theta'_{\rm{p}})
                       {\rm e}^{i(\phi_{\rm{\Lambda(1520)}} - \phi'_{\rm{p}})}$ \\ 
$\frac12$ & $-\frac12$ &$-(H_1{\rm e}^{i\phi_{H_1}} \d^{\frac12 }_{ \frac12 \frac12 }(\theta_{\rm{\Lambda(1520)}})
                \d^{\frac32 }_{    \frac12  -\frac12 }(\theta'_{\rm{p}}) {\rm e}^{i\phi'_{\rm{p}}} +
                       H_2{\rm e}^{i\phi_{H_2}} \d^{\frac12 }_{ \frac12    -\frac12 }(\theta_{\rm{\Lambda(1520)}})
                \d^{\frac32 }_{   -\frac12  -\frac12 }(\theta'_{\rm{p}}) {\rm e}^{i\phi_{\rm{\Lambda(1520)}}})$ \\ 
$-\frac12$ &  $\frac12$ & $ H_1{\rm e}^{i\phi_{H_1}} \d^{\frac12 }_{-\frac12   \frac12 }(\theta_{\rm{\Lambda(1520)}})
                ^{\frac32 }_{    \frac12   \frac12 }(\theta'_{\rm{p}}) {\rm e}^{-i\phi_{\rm{\Lambda(1520)}}} +
                       H_2{\rm e}^{i\phi_{H_2}} \d^{\frac12 }_{-\frac12    -\frac12 }(\theta_{\rm{\Lambda(1520)}})
                \d^{\frac32 }_{   -\frac12  \frac12 }(\theta'_{\rm{p}}) {\rm e}^{-i\phi'_{\rm{p}}} $ \\ 
$-\frac12$ & $-\frac12$ &$-(H_1{\rm e}^{i\phi_{H_1}} \d^{\frac12 }_{-\frac12 \frac12 }(\theta_{\rm{\Lambda(1520)}})
                \d^{\frac32 }_{    \frac12  -\frac12 }(\theta'_{\rm{p}}) 
                       {\rm e}^{-i(\phi_{\rm{\Lambda(1520)}} - \phi'_{\rm{p}})} +
                       H_2{\rm e}^{i\phi_{H_2}} \d^{\frac12 }_{-\frac12    -\frac12 }(\theta_{\rm{\Lambda(1520)}})
                \d^{\frac32 }_{   -\frac12  -\frac12 }(\theta'_{\rm{p}}))$ \\ 
\hline
\end{tabular}
\end{table}

\begin{table}[htbp]
\renewcommand{\arraystretch}{2.95}
\vskip 15pt
\caption{Amplitudes for the nonresonant $\Lambda_c^+ (\frac12^+) \to
pK^-\pi^+$ decay mode.}
\vskip 5pt
\label{amp4}
\begin{tabular}{ccc}\hline
$\rm{m}$ & $\lambda_{\rm{p}}$ & Amplitude \\ \hline
$ \frac12$ & $ \frac12$ & $N_{++}{\rm e}^{i\phi_{N_{++}}}$ \\ 
$ \frac12$ & $-\frac12$ & $N_{+-}{\rm e}^{i\phi_{N_{+-}}}$ \\ 
$-\frac12$ & $ \frac12$ & $N_{-+}{\rm e}^{i\phi_{N_{-+}}}$ \\ 
$-\frac12$ & $-\frac12$ & $N_{--}{\rm e}^{i\phi_{N_{--}}}$ \\ 
\hline
\end{tabular}
\end{table}

\section{Experiment E791 and Data Selection}

We analyze data from Fermilab fixed-target experiment E791, which ran
during 1991 and 1992.  The data were 
recorded from 500 GeV/$c$ $\rm{\pi^-}$ beam 
interactions in five thin target foils (one platinum, four diamond)
whose centers were separated 
by about 1.53 cm. The detector, described 
elsewhere in more detail\cite{exp:e791,e791_pairs}, was a large-acceptance, forward, 
two-magnet spectrometer.
The key components for this study were eight planes of multiwire 
proportional chambers, and six
planes of silicon microstrip detectors (SMD) before the target for 
beam tracking,
a 17-plane~SMD system and 35~drift chamber planes downstream of the target  
for track and vertex reconstruction,
and two multicell threshold {\v C}erenkov counters for charged particle
identification.

An unrestrictive, open-charm event selection based on total transverse
energy seen in the calorimeters was made in real time.
Offline we require all $\Lambda_c$ event candidates
to have a production and decay vertex longitudinally separated by at
least 6 $\sigma_{\rm{l}}$, where $\sigma_{\rm{l}}$ 
is the error in that separation. 
The tracks and vertex fits satisfy 
$\chi^2$ requirements. The vertex must be formed from proton,
kaon, and pion tracks identified as such by the {\v C}erenkov
particle identification system. 
Proton candidates are rejected when projected into
regions of the detector with poor particle identification efficiency.
To reject more background, we require that the magnitude of the
vector sum of the transverse momenta of all the
secondary tracks with respect to the flight path of the reconstructed
$\Lambda_c$ candidate be $\leq$ 0.4 GeV/$c$. To eliminate
reflections from
$\rm{D^+\to K^-\pi^+\pi^+}$ decays and from $\rm{D^+}$, 
$\rm{D_s^+\to K^-K^+\pi^+}$
decays, we remove all events whose reconstructed $\rm{K\pi\pi}$ mass
is within the range [1.85, 1.89] GeV/$c^2$ or 
whose reconstructed $\rm{KK\pi}$ mass is within the
ranges [1.85, 1.89] GeV/$c^2$ or [1.95, 1.99] GeV/$c^2$.

After this preliminary stage of the analysis, we have 998$\pm$167 
$\Lambda_c^+\to pK^-\pi^+$ signal events and 
107\,368$\pm$366 background events.  Therefore we apply further
requirements in order to improve this signal in an unbiased way. 
This is accomplished by a selection on 
the output of an artificial neural network. The network is trained
using a Monte Carlo (MC) sample of $\Lambda_c^+\to pK^-\pi^+$ decays 
for signal and events from the wings of the $pK^-\pi^+$ 
mass distribution in data for background. The variables used in the
training include all those described in the preliminary analysis above
and the transverse miss distance between the primary vertex and the 
line of flight of the reconstructed $\Lambda_c$ candidate, the scalar 
sum of the $p_T^2$ of all the
secondary tracks with respect 
to the flight path of the reconstructed $\Lambda_c$ candidate, 
the significance of the separation between the secondary
vertex and the closest target foil edge,
the calculated proper lifetime of the reconstructed $\Lambda_c$,
the ratio of the distance of each of the decay tracks from the secondary
vertex to its distance from the primary vertex, and the 
minimum and product of the three ratios.

\begin{figure}[htbp]
        \centering \leavevmode
        \epsfxsize=3.0in
        \epsfbox[20 150 530 650]{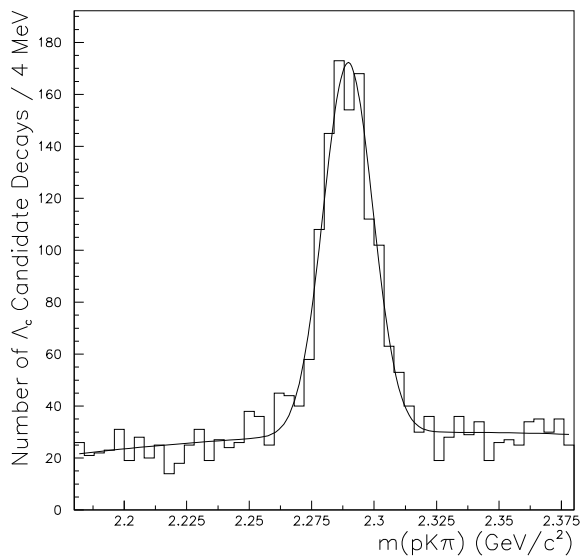}
        \caption{$\Lambda_c\rightarrow pK\pi$ signal used in this
                 analysis.}
\label{postnncut}
\end{figure}

The cut on the neural
net output is chosen to maximize 
$N_{\mathrm{S,MC}} / \sqrt{N_{\mathrm{S,MC}} + N_{\mathrm{B}}}$ 
where $N_{\mathrm{S,MC}}$
is the number of MC signal events (scaled to our data size) and $N_B$
the number of background events in the signal region.
After the cut on the neural net
output, 2271 real events in the $pK^-\pi^+$ mass range of 
[2.18, 2.38] GeV/$c^2$ survive, as seen in Figure \ref{postnncut}.

\section{Differential Decay Rate Fit}

Once the final data set is established we constrain each reconstructed
$pK\pi$ mass to the mass of the $\Lambda_c$. This forces
all candidate events, whether in the signal peak or in the wings, 
to have the same decay phase space as that of the signal. We
describe each event by the five independent kinematic
variables of interest (m$^2$($K\pi$), m$^2$($p\pi$),
cos($\theta_{\rm{p}}$), $\phi_{\rm{p}}$, 
and $\phi_{K\pi}$). 

Next we use an extended maximum likelihood technique to fit the data. 
The likelihood is assumed to have the form
        \begin{equation}
        \ln{{\cal L}} =  \sum_i \ln{{\cal L}_i}
        - \left({1\over2}\ln{[2\pi N_{\rm{pred}}]}
        + {[N_{\mathrm{pred}} 
	- N_{\mathrm{obs}}]^2\over 2 N_{\mathrm{pred}}}\right),
        \label{eqlike}
        \end{equation}
where $N_{\mathrm{pred}} = N_{\mathrm{S}} + N_{\mathrm{B}}$ and 
$N_{\mathrm{obs}}$ are the predicted and observed numbers of
events, and ${\cal L}_i$ is the likelihood of each event defined by
a joint probability density in the five-dimensional phase space of the decay
and the one-dimensional space of unconstrained $m_{pK\pi}$ 
(Fig. \ref{postnncut}).

The likelihood for an individual event is given by 
\begin{equation}
{\cal L}_i = {N_S \over N_S + N_B} G(m_{pK\pi,i},x_{\rm{F}}) 
						S(\vec x_i) A(\vec x_i)
           + {N_B \over N_S + N_B} Q(m_{pK\pi,i}) B(\vec x_i)
\label{eqlsubi}
\end{equation}
Here $G$ and $Q$ represent normalized Gaussian and quadratic functions of
the three body mass $m_{pK\pi}$ while $S(\vec x)$, $A(\vec x)$, 
and $B(\vec x)$ are the signal, acceptance, 
and background normalized densities in
the five-dimensional space of decay kinematic variables described
previously. Note that the width of $G$ is dependent on $x_{\rm{F}}$, 
the scaled longitudinal momentum of the $\Lambda_c$-candidate, and is 
parameterized as $\sigma = \sigma_a\times x_{\rm{F}} + \sigma_b$. Our
$x_{\rm{F}}$ range is essentially [-0.1, +0.4].
The function $S(\vec x)$ is described 
in Equation \ref{eqsigdens}, while $A(\vec x)$ is
a model of the acceptance, a five-dimensional density of the reconstructed 
and surviving uniformly-generated MC events, and $B(\vec x)$ is a
model of the background, coming from a five-dimensional 
density of data events in the wings of the
mass plot. We model our acceptance, $A(\vec x)$, and background, 
$B(\vec x)$,  using the same approach. This common
procedure is based on the $K$-nearest-neighbors method\cite{bish}.  
The $K$-nearest-neighbors method works on the 
principle that the density at a point in phase space 
is proportional to $N/V(r(\vec{x}_i))$ where $N$ is 
the number of events which occur within volume $V(r(\vec{x}_i))$ which
is a multidimensional sphere of radius $r$ centered at $\vec{x}_i$.
This method is augmented to ensure that each of the one-dimensional
projections of the five-dimensional density in question ($A(\vec x)$ 
and $B(\vec x)$) matches the corresponding projection of the data set
modeled (accepted MC events and background events from the data wings,
respectively).

\begin{figure}[htbp]
        \centering \leavevmode
        \epsfxsize=3.0in
        \epsfbox[18 157 528 656]{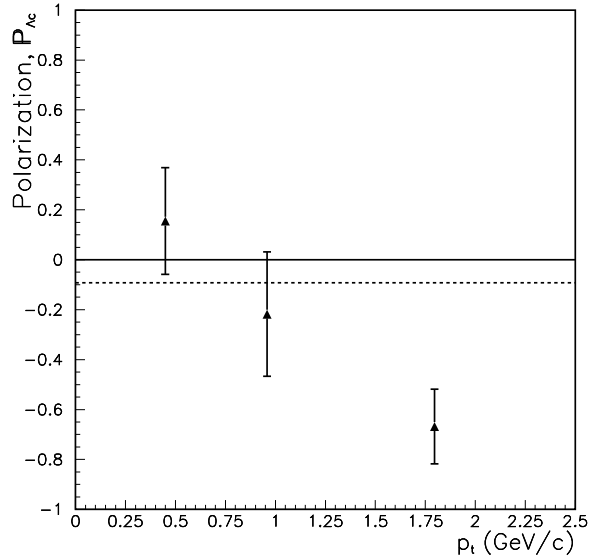}
        \caption{The polarization of the $\Lambda_c$ as a function of the
$\Lambda_c$'s transverse momentum. The vertical bars represent the
error as found by MINUIT.  They are placed at the average $p_T^2$
value for that region. Bin sizes are as in Table~\protect\ref{fit3}. 
The dotted line is the value of the polarization integrated over
$p_T^2$.}
\label{ppt}
\end{figure}

Using the fitting software, MINUIT\cite{min}, we find the parameter values 
which maximize $\cal L$ in Equation \ref{eqlike}.  The results 
can be seen in Table \ref{fit3}.  
Because the polarization is dependent on transverse momentum\cite{dha}, 
we split the data set into three equally populated 
regions of $p_T^2$ (0.00--0.70 GeV/$c$, 0.70--1.24 GeV/$c$, and 1.24--5.20 GeV/$c$)
and find the polarization for each region (${\bf P}_{\Lambda_c,1}$,
${\bf P}_{\Lambda_c,2}$, and ${\bf P}_{\Lambda_c,3}$, respectively).
The polarization versus $p_T^2$ is shown in Figure \ref{ppt}.

\begin{table}[hbtp]
\renewcommand{\arraystretch}{1.10}
\caption{Decay amplitudes, polarization, and mass plot parameters for
  $\Lambda_c\rightarrow pK\pi$ from the MINUIT fit with statistical
  errors. The two parameters without errors are the only ones that are
  fixed.
}
\vskip 3pt
\label{fit3}
{\SS
\begin{tabular}{ccc}\hline
Terms & Parameter & Value \\ \hline
$p\overline{K}^{*0}(890)$ & $E_1$ &                  
    0.52$\pm$    0.17 \\
 & $\phi_{E_1}$ &                               
   -1.01$\pm$    0.48 \\
 & $E_2$      &                                 
    0.20$\pm$    0.10 \\
 & $\phi_{E_2}$ &                               
    2.35$\pm$    0.67 \\
 & $E_3$      &                                 
    0.21$\pm$    0.10 \\
 & $\phi_{E_3}$ &                               
    3.46$\pm$    0.42 \\
 & $E_4$      &                                 
    0.16$\pm$    0.10 \\
 & $\phi_{E_4}$ &                               
    5.29$\pm$    0.55 \\
  
$\Delta^{++}(1232)K^-$ & $F_1$ &                      
    0.17$\pm$    0.07 \\
 & $\phi_{F_1}$ &                               
    4.98$\pm$    0.41 \\
 & $F_2$      &                                 
    0.38$\pm$    0.13 \\
 & $\phi_{F_2}$ &                               
    4.88$\pm$    0.40 \\
  
$\Lambda^*(1520)\pi^+$ & $H_1$ &              
    0.18$\pm$    0.09 \\
 & $\phi_{H_1}$ &                               
    5.93$\pm$    0.52 \\
 & $H_2$      &                                 
    0.20$\pm$    0.07 \\
 & $\phi_{H_2}$ &                               
   -0.06$\pm$    0.55 \\
  
Nonresonant & N$_{++}$ &                        
    0.46$\pm$    0.26 \\
 & $\phi_{N_{++}}$ &                            
    3.48$\pm$    0.54 \\
 & N$_{+-}$ &                                   
    1.00 \\
 & $\phi_{N_{+-}}$ &                            
    0.00 \\
 & N$_{-+}$ &                                   
    0.18$\pm$    0.15 \\
 & $\phi_{N_{-+}}$ &                            
    0.75$\pm$    0.71 \\
 & N$_{--}$ &                                   
    0.94$\pm$    0.45 \\
 & $\phi_{N_{--}}$ &                            
    1.13$\pm$    0.36 \\
  
Polarization [0 $ < p_T^2 < $ 0.70 GeV/$c^2$] & ${\bf P}_{\Lambda_c,1}$ &        
    0.15$\pm$    0.21 \\
  
Polarization [0.70 $ < p_T^2 < $ 1.24 GeV/$c^2$] & ${\bf P}_{\Lambda_c,2}$ &        
   -0.22$\pm$    0.25 \\
  
Polarization [1.24 $ < p_T^2 < $ 5.20 GeV/$c^2$] & ${\bf P}_{\Lambda_c,3}$ &        
   -0.67$\pm$    0.15 \\
  
\# Signal Events & $N_S$ &                       
  946$\pm$   38 \\
  
\# Background Events & $N_B$ &                        
 1324$\pm$   43 \\
  
Background Quadratic Term & $b_q$ &                          
   -0.98$\pm$   10.51 \\
  
Background Linear Term & $b_l$ &                        
    1.34$\pm$    0.48 \\
  
$\Lambda_c$ Mass (GeV/$c^2$) & m$_0$ &        
    2.29$\pm$    0.00 \\
  
$\Lambda_c$ Width (MeV/$c^2$) & $\sigma_a$ &  
   20.1$\pm$    4.8 \\
\omit & $\sigma_b$ \vspace{1mm} &  
    9.3$\pm$    0.6 \\
  \hline
\end{tabular}}
\end{table}

\begin{figure}[hbtp]
        \centering \leavevmode
        \epsfxsize=5.0in
        \epsfbox[71 144 536 655]{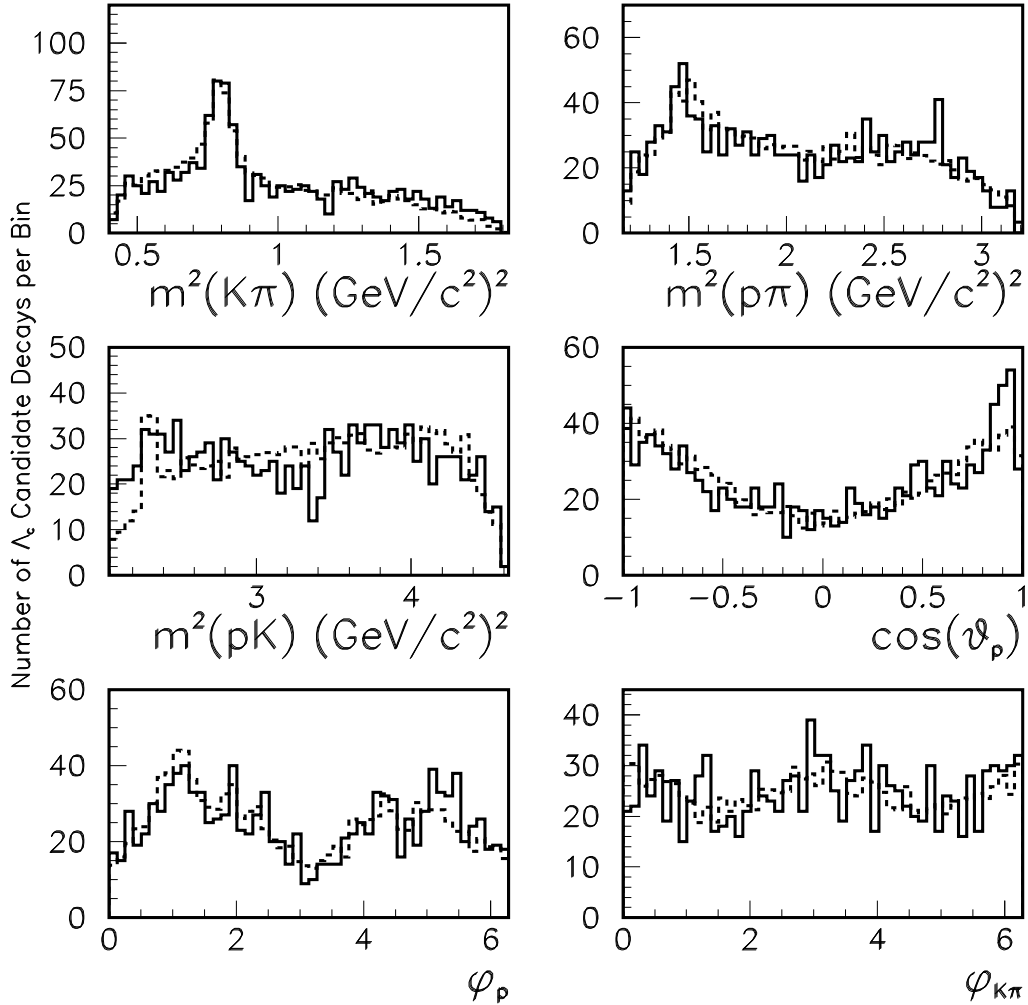}
        \caption{Projections of the data within the $m_{pK\pi}$
range of (2.265 -- 2.315 GeV/$c^2$) and overall fit
superimposed. The data are shown as a solid histogram and the fits as
dashed lines. There are 50 bins in each plot.} 
\label{c_finalfit_xfpt_31_sr_04_19n_42rv}
\end{figure}

\begin{figure}[hbtp]
        \centering \leavevmode
        \epsfxsize=5.0in
	\epsfysize=8.0in
        \epsfbox[71 144 536 655]{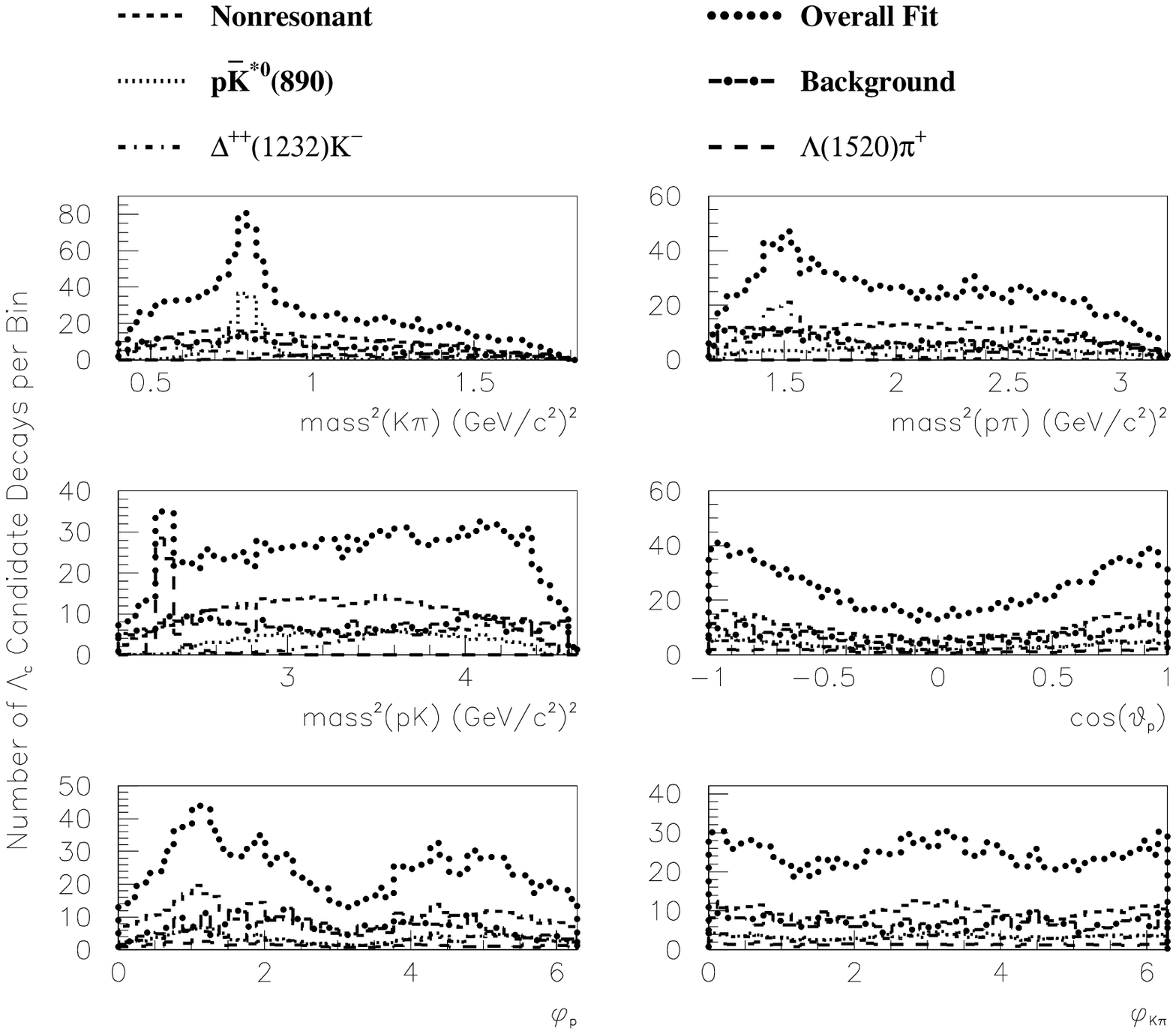}
        \caption{Projections of the overall fit components scaled to
match the number of data events within the m($pK\pi$) range of
(2.265 -- 2.315 GeV/$c^2$). There are 50 bins in each plot.}
\label{cff_xfpt_31_sr_04_19n_42rv}
\end{figure}

Figure \ref{c_finalfit_xfpt_31_sr_04_19n_42rv} shows the 
one-dimensional projections of the overall fit using
the parameters of Table \ref{fit3} and the data in the signal region.  The
$\chi^2$/DOF for the agreement of the one-dimensional 
projections of the model and the data is 1.06.
Although this agreement is reasonable, there are discrepancies, especially
in the pK-mass-squared distribution.  A model with a spin $\frac12^-$ resonance
decaying to pK and having a mass of 1.556$\pm$0.019 GeV/$c^2$ and width of 
279$\pm$74 MeV/$c^2$
reduces the overall $\chi^2$ of the fit.  However, there is no known
resonance with this description.  Trying another fit model, we find 
some evidence for an upper tail of the $\Lambda(1405)(\to pK)$ \cite{dal98}.
Similarly, an improvement may be obtained by
inclusion of nonresonant contributions beyond s-wave.
In the case of each of these other models, the data are not
sufficient to provide compelling evidence and/or to give significant
measurements of the required additional parameters for the 
models.  Figure \ref{cff_xfpt_31_sr_04_19n_42rv} shows
the one-dimensional projections of the 
model's resonant and nonresonant components, background, and total fit.

We calculate the fit fraction associated 
with each decay using Equation \ref{ff}.
A fit fraction $F_i$ is the fraction of our signal described
by decay via the $i^{th}$ contribution and is the ratio
of the magnitude squared of the amplitude term describing this contribution
and the complete probability of the decay:
\begin{equation}
F_i = {\int \sum_{  m,\lambda_{\rm{p}}} | B_i(m_i)\alpha_{i,m,\lambda_{\rm{p}}} |^2 \,\d\vec{x}\over 
       \int \sum_{m,\lambda_{\rm{p}}} | \sum_j B_j(m_j)\alpha_{j,m,\lambda_{\rm{p}}} |^2\,\d\vec{x}}
\label{ff}
\end{equation}
using the notation of Equation \ref{eqsigdens} where $\int \d\vec{x}$ implies integrating over the decay phase space.

The major systematic errors on the fit fractions come from uncertainties in the performance of
our \v{C}erenkov counters and drift chambers, 
in the MC model for the production of $\Lambda_c^+$ baryons, and 
in the adjustment of the 
$K$-nearest-neighbors model to the one-dimensional projections. 
In each case, a different model
of acceptance is studied and the change in the central values of
parameters is taken to be an estimate of the systematic error. Table
\ref{systerr} lists the uncertainties in the fit fractions due to these
various sources. 

\begin{table}[htbp]
\caption{The systematic errors on fit fractions.}\label{systerr}
\vskip 3pt
\begin{tabular}{cccccc}\hline
     & Drift & {\v C}erenkov & Acceptance & Production & Combined \\
Mode & Chamber ($\%$) &  Counter ($\%$) & Adjust ($\%$) & Model ($\%$) & Error ($\%$) \\ \hline
$p\overline{K}^{*0}(890)$ &    1.7 & 0.6 & 0.3 & 0.0 & 1.8 \\ 
$\Delta^{++}(1232)K^-$     &    2.5 & 1.2 & 0.7 & 0.1 & 2.9 \\ 
$\Lambda(1520)\pi^+$ &  0.9 & 0.5 & 0.4 & 0.1 & 1.1 \\ 
Nonresonant          	  &    3.2 & 1.1 & 0.9 & 0.1 & 3.5 \\ \hline
\end{tabular}
\end{table}

The fit fractions for $\Lambda_c^+\rightarrow pK^-\pi^+$ 
are listed in Table \ref{fitfracfinal}, 
along with the statistical (first) and systematic (second) uncertainties.  
A check of these results is made by fitting the data 
assuming a single, average polarization.  
The resulting polarization is ${\bf P}_{\Lambda_c}$
= -0.09$\pm$0.14 (shown as the dotted line in Figure \ref{ppt}).  
The component fit fractions
do not change significantly.

\begin{table}[htbp]
\caption{The fit fractions for $\Lambda_c^+\rightarrow pK^-\pi^+$
with statistical and systematic errors from the final fit.}
\vskip 3pt
\label{fitfracfinal}
\begin{tabular}{cc}\hline
Mode & Fit Fraction ($\%$) \\ \hline
$p\overline{K}^{*0}(890)$	& 19.5$\pm$2.6$\pm$1.8 \\ 
$\Delta^{++}(1232)K^-$		& 18.0$\pm$2.9$\pm$2.9 \\ 
$\Lambda(1520)\pi^+$	&  7.7$\pm$1.8$\pm$1.1 \\ 
Nonresonant         		& 54.8$\pm$5.5$\pm$3.5 \\ \hline
\end{tabular}
\end{table}

Our results are compared to other published results in Table \ref{other}.
The components in this decay do not demonstrate significant interference.
This is most easily demonstrated by the fact that the fit
fractions in Table \ref{fitfracfinal} sum to near unity.  
Although we have better statistics than previous measurements, 
the uncertainties are comparable due to our more general fit.
\begin{table}[htbp]
\caption{$\Lambda_c$ branching ratios relative to the inclusive 
$\Lambda_c^+\rightarrow pK^-\pi^+$ branching fraction.  
The NA32 and ISR values were calculated from one-dimensional projections
only.}
\vskip 3pt
\label{other}
\begin{tabular}{cccc}\hline
Mode & E791 & NA32\cite{boz} & ISR\cite{bas} \\ \hline
$p\overline{K}^{*0}(890)$    & 0.29$\pm0.04\pm$0.03 & 0.35$^{+0.06}_{-0.07}\pm$0.03 & 0.42$\pm$0.24 \\  
$\Delta^{++}(1232)K^-$   & 0.18$\pm0.03\pm$0.03 & 0.12$^{+0.04}_{-0.05}\pm$0.05 & 0.40$\pm$0.17 \\ 
$\Lambda(1520)\pi$ & 0.15$\pm0.04\pm$0.02 & 0.09$^{+0.04}_{-0.03}\pm$0.02 & \\ 
Nonresonant 	   	& 0.55$\pm0.06\pm$0.04 & 0.56$^{+0.07}_{-0.09}\pm$0.05 & \\ \hline
\end{tabular}
\end{table}

\section{Discussion and Conclusions}

Significant resonant and non-resonant branching fractions are found
in this analysis of $946 \pm 38$ $\Lambda_c^+\rightarrow pK^-\pi^+$
signal events, the largest sample ever
analyzed.  The size of the sample allows for inclusion of relative 
phases of the various contributions and a full accounting of spin and
production polarization for the first time in such an analysis.   The
$\Delta(1232)^{++}K^-$ and $\Lambda(1520)\pi^+$ decay modes 
are seen as statistically
significant contributions for the first time, even when uncertainties
associated with phases and other variables are included.  
The observation of a substantial $\Delta^{++}K^-$ component 
provides strong evidence for the 
W-exchange amplitude in charm baryon decays. The observed components of the
$\Lambda_c^+\rightarrow pK^-\pi^+$ decay do not 
interfere significantly.  Finally, we find no evidence for either 
$\Lambda(1600)\pi^+$ or $\Sigma(1660)\pi^+$ in the data.

In spite of the improvements in the $\Lambda_c^+\rightarrow
pK^-\pi^+$ results coming from inclusion of the newly observed
contributions, there remains poor agreement between the fit and the data 
in the pK-mass-squared projection. Additional data from new
experiments are needed in order to conclusively demonstrate additional
resonances (or their tails).

We find evidence for an increasingly negative polarization of 
the $\Lambda_c$ baryons as a function of
$p_T^2$, in agreement with a recent model\cite{dha} and, for large
negative decay asymmetry, the measurement of Ref. \cite{jez}. As suggested
by the authors of the model, the assumption of T-invariance leads to
further constraints on resonant phases. Again, our data are consistent
with these constraints but additional data from new
experiments are needed for a stringent test of CP violation.

\section{Acknowledgements}

It is a pleasure to acknowledge useful conversations with
William Dunwoodie, Gary Goldstein, and Fred Myhrer.
We gratefully acknowledge the staffs of Fermilab and of all the
participating institutions. This research was supported by the Brazilian
Conselho Nacional de Desenvolvimento Cient\'{\i}fico e Technol\'ogio, the
Mexican Consejo Nacional de Ciencia y Tecnologica, the Israeli Academy
of Sciences and Humanities, the U.S. Department of Energy, the
U.S.-Israel Binational Science Foundation, and the U.S. National Science
Foundation. Fermilab is operated by the Universities Research
Association, Inc., under contract with the United States Department of
Energy.

\end{document}